\documentclass[fleqn,10pt]{wlscirep}
\usepackage[utf8]{inputenc}
\usepackage[T1]{fontenc}
\usepackage{lineno}
\usepackage{amsmath}

\title{Testing NeuralGCM's capability to simulate future heatwaves based on the 2021 Pacific Northwest heatwave event}

\author[1,*]{Shiheng Duan}
\author[1]{Jishi Zhang}
\author[1]{Céline Bonfils}
\author[1]{Giuliana Pallotta}

\affil[1]{Lawrence Livermore National Laboratory}

\affil[*]{duan5@llnl.gov}

\begin{abstract}
AI-based weather and climate models are emerging as accurate and computationally efficient tools. Beyond weather forecasting, they also show promise to accelerate storyline analyses. We evaluate NeuralGCM's ability to simulate an extreme heatwave against the Energy Exascale Earth System Model (E3SM), a physics-based climate model. NeuralGCM accurately replicates the targeted event, and generates stable and realistic mid-century projections. However, due to the absence of land feedbacks, NeuralGCM underestimates the projected warming amplitude compared to physics-based model references. 
\end{abstract} 
\begin{document}

\flushbottom
\maketitle

\thispagestyle{empty}


Extreme heatwaves, such as the unexpected 2021 Pacific Northwest heatwave, are projected to become more frequent and severe due to climate change, posing significant risks to human health, ecosystems, and infrastructure \cite{IPCC_2021_WGI_Ch_11}. Understanding how these heatwave events (or any other type of extreme weather events) will evolve in response to anthropogenic climate change is critical for effective risk assessment and adaptation strategies. Among approaches targeting this goal, storyline analyses explicitly aim at quantifying how past weather events would unravel under alternative conditions (e.g., different climate forcing scenarios or adaptation measures)\cite{shepherd2018storylines}. A prominent approach is the pseudo-global warming (PGW) simulation, which is grounded on the premise that synoptic dynamical conditions should remain unchanged to preserve the sequence of historical weather events, while modified thermodynamic signals are imposed to assess the impact from alternative warming conditions. Such simulations help answer the question of how a specific extreme event would have unfolded in a warmer climate. PGW analyses conducted with physics-based models typically rely on nudging techniques to adjust the dynamical fields to observations and ensure that the simulations reproduce the atmospheric circulation patterns occurring during targeted events \cite{Zhang_2024}. Compared to simulations in which dynamic and thermodynamic fields intertwine and freely evolve, the nudging approach enables a clearer framework for analyzing the factors contributing to extreme events under different climate scenarios \cite{risser2024granger}. 

The emergence of Artificial Intelligence (AI) has enabled the development of computationally efficient data-driven weather models \cite{eyring2024pushing}. Although their focus has been on improving weather forecast, their applications in storyline analyses holds great promise. Several key limitations exist, however: first, most AI-based weather models struggle to extrapolate into different climate conditions. As shown by a reference study\cite{rackow2024robustness}, most models have not been designed for climate extrapolation and show instability in this regard, regressing towards their training climatology. Second, unlike physics-based models, it is a non-trivial task to directly implement nudging with current AI-based models. Indeed, there are concurrent studies focusing on data assimilation, as relevant a topic as nudging, showing issues with stability and accuracy \cite{slivinski2024assimilating}. Finally, these atmospheric-only AI models do not include land or ocean components, which hinders feedbacks from these domains. 

In this study, we jointly address all these challenges and select NeuralGCM \cite{kochkov2024neural}, a cutting-edge hybrid model, as our AI-based model to evaluate though the lens of storyline analysis. NeuralGCM is trained on ERA5 reanalysis dataset\cite{hersbach2020era5} and has proven to be stable across several decades. At the $1.4^\circ$ resolution used in this study, it is available in two forms: a deterministic version and a stochastic version. Since the current model version lacks explicit nudging capabilities, we introduce two methods to constrain dynamical conditions. The first one follows the nudging strategy in physical-based models by calculating the difference between simulated and reference variables, and applying the nudging tendency based on a predefined relaxation timescale. This is applied to the deterministic version of NeuralGCM and referred to as ``NeuralGCM-Nudge" (abbreviated to ``NGCM-Nudge"). The second proposed method, \textit{ensemble involution}, is a novel way to constrain dynamical conditions towards the observed state by iteratively comparing and discarding ensemble members whose dynamical fields deviate from the target. Although technically distinct from the traditional nudging, both methods share the same fundamental purpose of constraining dynamical conditions in PGW simulations. Since this second method relies on ensemble simulations, it applies to the stochastic version and its results are referred to as ``NeuralGCM-ENS” (NGCM-ENS). The implementation details are available in the Method section. 

As our testbed to evaluate NeuralGCM in storyline configuration, we choose the June 26th-July 2nd 2021 Pacific Northwest heatwave, an unprecedented weather event showing record-high temperatures which is not included in the NeuralGCM's training dataset. We first evaluate NeuralGCM's ability to replicate this extreme event compared to observations and hindcasts of the event performed with the Energy Exascale Earth System Model (E3SM) \cite{golaz2022doe}. We then evaluate how this event would unravel under warmer conditions using NeuralGCM, and compare these results against E3SM storyline simulations. Here,  E3SM simulations are conducted using two land spin-up configurations (``current" and ``future") to account for different land feedback effects (Methods).

\begin{figure}[th!]
\centering
  \includegraphics[width=.9\linewidth]{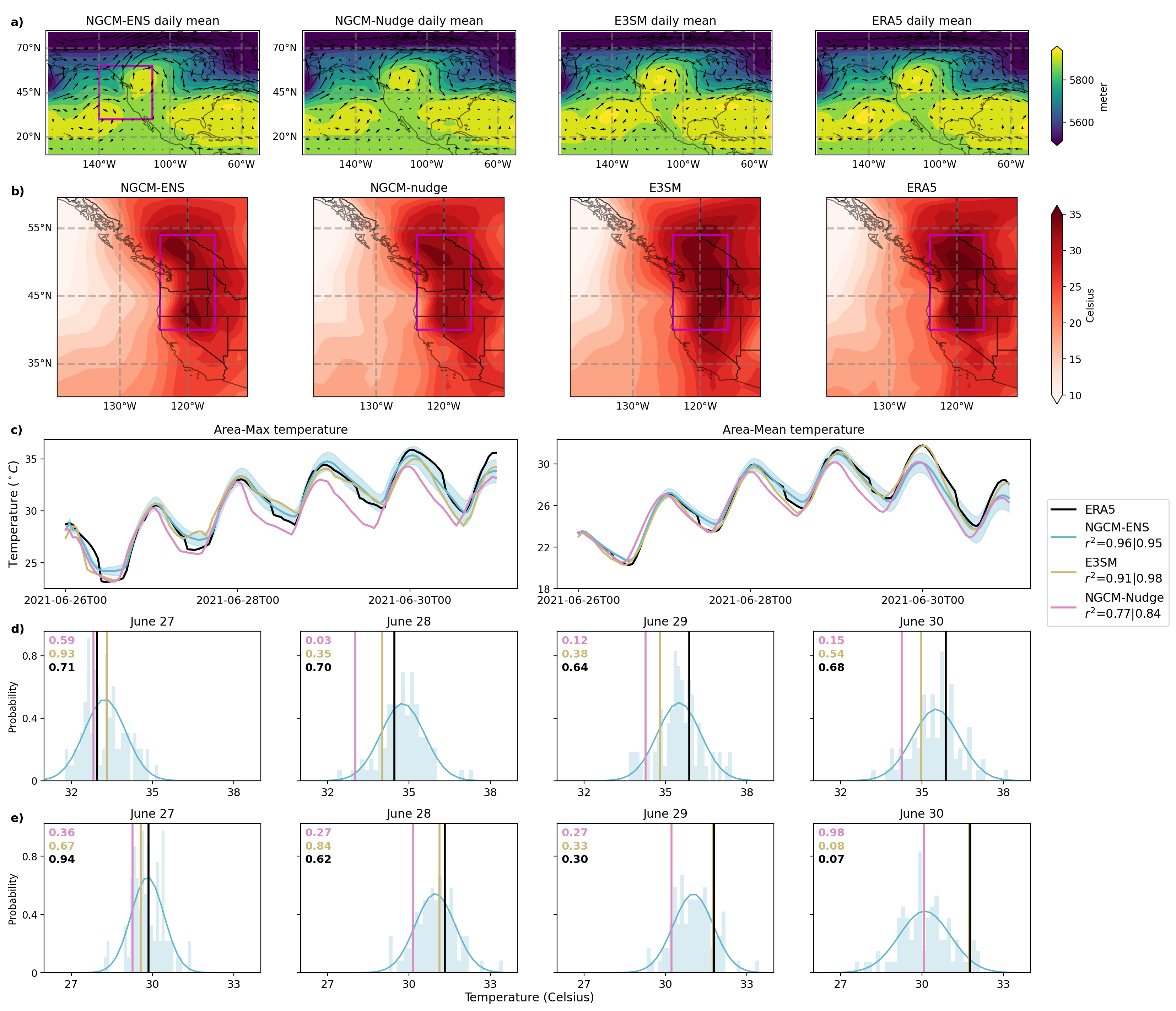}
  \caption{Hindcast results from NeuralGCM and E3SM, compared to ERA5 reference for \textbf{a)} daily average large scale circulations (500 hPa geopotential height and wind vectors) and \textbf{b)} daily maximum 850 hPa temperature based on hourly averages on 2021 June 30th. Ensemble mean is used for NeuralGCM (stochastic version), and denoted as ``NGCM-ENS". ``NGCM-Nudge" represents the nudging results from NeuralGCM (deterministic version). The magenta square in \textbf{a)} NGCM-ENS denotes the ensemble involution domain, and in \textbf{b)} the heatwave core region. \textbf{c)} Hourly area-maximum (left) and area-mean (right) temperature profiles in the heatwave core region. The ERA5 reference is displayed in black. The blue line represents NeuralGCM ensemble mean (NGCM-ENS), with the shaded area illustrating the ensemble variability (ensemble mean $\pm$ one standard deviation). The pink and yellow lines display nudging results from NGCM-Nudge and E3SM, respectively. Coefficients of determination between ERA5 and the NGCM/E3SM results are reported in the legend, with the first value representing area-max and the second for area-mean. Daily hottest \textbf{d)} area-maximum and \textbf{e)} area-mean temperature are examined over the core region. In \textbf{d)} and \textbf{e)}, the histograms of NGCM-ENS temperature are approximated via a Normal distribution (solid blue line) , the black vertical line represents ERA5 temperature, while the pink and yellow vertical lines correspond to the nudging results for NGCM-Nudge and E3SM, respectively. In each panel, the two-sided p-values indicate the probability of observing NGCM-Nudge, E3SM, and ERA5 temperature values as extreme or more extreme than the temperature range encompassed under the NGCM-ENS distribution. These p-values are color-coded for clarity. 
  }
  \label{fig:hindcast}
\end{figure}

We first examine the E3SM- and NeuralGCM-based hindcasts (Fig.\ref{fig:hindcast}). Since surface temperature is not directly available in NeuralGCM, 850hPa temperature is used as the target variable. As expected, both NeuralGCM-Nudge and E3SM successfully reproduce the atmospheric dynamics by nudging the ERA5 wind fields, accurately simulating the high-pressure system over the Pacific Northwest, a key factor for this heatwave. NeuralGCM-ENS (ensemble mean) also captures a comparable high-pressure system, demonstrating the effectiveness of our ensemble involution method, which considerably reduced our ensemble size from the initial 1000 to 86 members, while constraining the dynamical conditions. In addition to the daily circulation examined in Fig.\ref{fig:hindcast}, an evaluation of hourly wind profiles is available in Fig.S1. The results show comparable agreement with ERA5 across NeuralGCM-ENS, NeuralGCM-Nudge and E3SM, with slightly reduced performance for NeuralGCM-ENS at 500 hPa. These results further support the effectiveness of nudging and ensemble involution for NeuralGCM. However, subtle differences in circulation exist. For example, the circulation near $60^\circ N$ $140^{\circ}W$ shows a cyclonic feature in both E3SM and NeuralGCM-Nudge stronger than NeuralGCM-ENS. This is simply due to the fact that the ensemble involution is carried out regionally, whereas nudging is applied globally (a sensitivity test on ensemble involution can be found in Method). On June 30th, the daily maximum temperature over the Pacific Northwest from NeuralGCM and E3SM (Fig.\ref{fig:hindcast}\textbf{b}) show similar spatial distributions as ERA5, and the extreme heat extends from British Columbia to North Nevada. Two temperature peaks are observed in the ERA5 reference over southern British Columbia and eastern Oregon. Both NeuralGCM-ENS and NeuralGCM-Nudge are able to reproduce these centers, though their intensity is underestimated (Fig.S2). In contrast, E3SM tends to overestimate the heatwave spatial extent, failing to capture distinct heat clusters with overestimation in eastern Washington. When zooming into the heatwave core region (magenta box in Fig.\ref{fig:hindcast}\textbf{b}), the temperature time-series (Fig.\ref{fig:hindcast}\textbf{c}) reveals that NeuralGCM-ENS and E3SM exhibit comparable performance in replicating ERA5-derived area-maximum and area-mean temperature evolution, with relatively high coefficients of determination. In contrast, NeuralGCM-Nudge systematically underestimates both metrics. Notably, the location of area-maximum temperature changes through time and varies among models. To gain deeper insights into NeuralGCM's performance, the mean absolute error and root mean squared error are calculated at each time step, followed by spatial averaging over the heatwave core area to derive an area-mean error time series(Fig.S2). We obtain that all models exhibit increasing error with time. NeuralGCM-Nudge generally has a slightly higher but comparable error against NeuralGCM-ENS, whereas E3SM shows a larger overall error with a pronounced diurnal cycle, characterized by greater discrepancies at night. 

Since NeuralGCM-ENS provides a distribution of simulations, we further evaluate it in a probabilistic manner.
For each day, the daily hottest area-maximum and area-mean temperatures from E3SM and NeuralGCM-Nudge are compared against the ensemble distribution, with the ERA5 values serving as reference (Fig.\ref{fig:hindcast}\textbf{d} and \textbf{e}). Overall, the ERA5 reference is statistically compatible to the NeuralGCM-ENS distribution, generally falling near its center. Under the null hypothesis that ERA5 temperatures are drawn from the same distribution as NeuralGCM-ENS, the p-values exceed the 0.05 significance threshold on all days, indicating insufficient evidence in all cases to reject the null hypothesis. Although the p-value is relatively lower in the June 30 / area-mean case (Fig.1e; 0.07), it still supports statistical consistency, suggesting that the ERA5 temperatures remain within the ensemble variability. 
Similar results are found for E3SM and NeuralGCM-Nudge, with only one exception on June 28th for the latter. This further indicates that nudging and ensemble involution are comparable methods for generating \textit{plausible} temperature outcomes. Notably, while E3SM can generate a heatwave with temperature either similar or lower than in ERA5, NeuralGCM-Nudge shows consistent lower temperature and is generally located to the left side of ERA5 (especially in the June 30 / area-mean case), as shown in Fig.\ref{fig:hindcast}\textbf{d} and \textbf{e}. However, this should not necessarily be interpreted as a model deficiency, as extreme events inherently involve unpredictability, and variations in initial conditions can lead to substantially different outcomes \cite{lerch2017forecaster}. 

\begin{figure}[ht]
\centering
  \includegraphics[width=.9\linewidth]{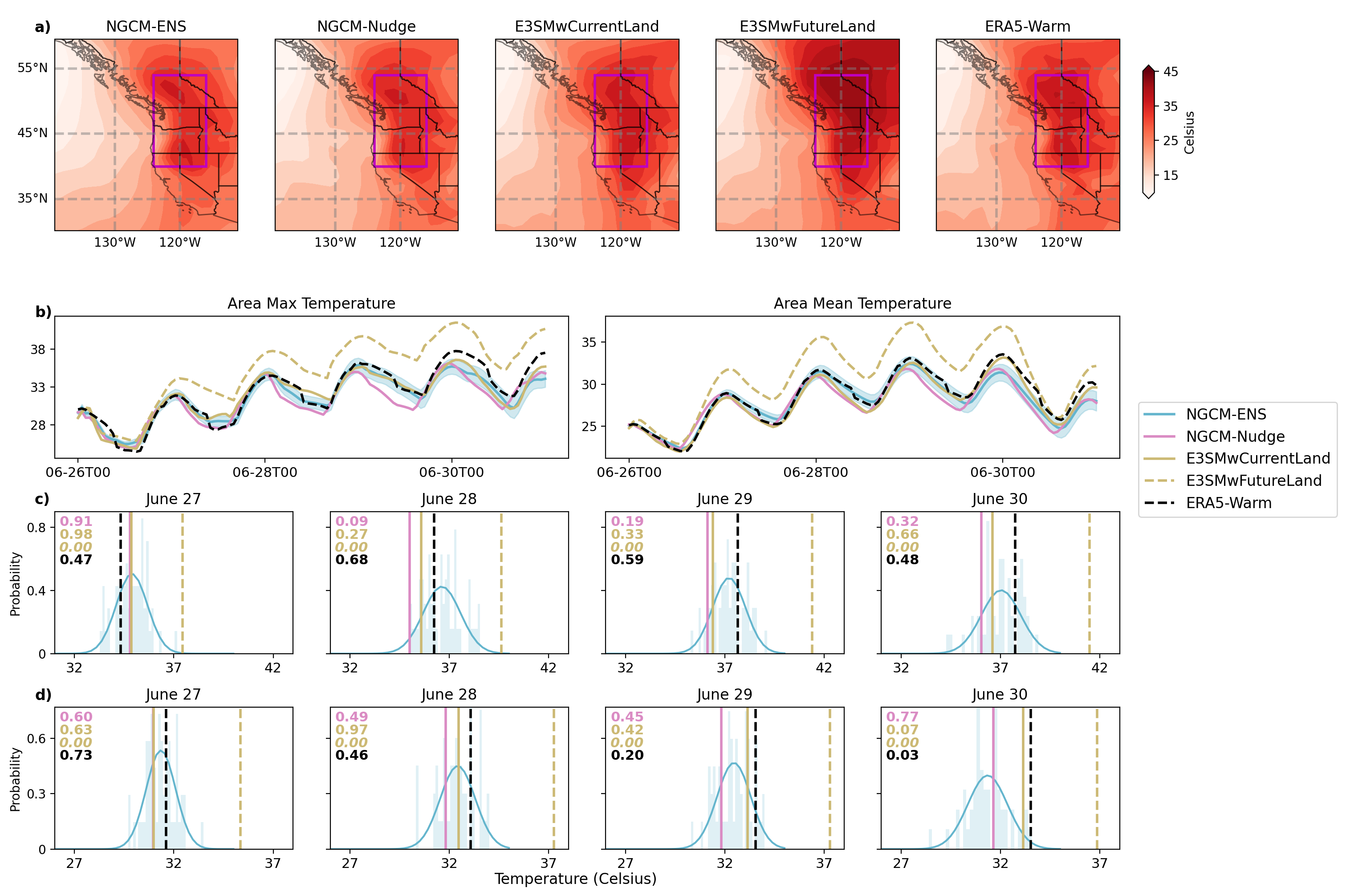}
  \caption{Mid-century storyline results from NeuralGCM and E3SM. \textbf{a)} Projection of daily maximum 850 hPa temperature (based on hourly timesteps) on June 30th (note the different temperature scale for the colorbar than Fig.1\textbf{b}). Ensemble mean (in this case the ensemble size was reduced from 1000 to 54 members) is used for NeuralGCM-ENS (abbreviated as ``NGCM-ENS"). The nudging results from NeuralGCM is labeled as ``NGCM-Nudge". Two land settings are used for E3SM. The E3SM with land spun-up by current forcings is denoted as ``E3SMwCurrentLand", while ``E3SMwFutureLand" refers to results where land is initialized with projected warming conditions. As a reference, ``ERA5-Warm" represents the case where the climate change delta signal is directly added to ERA5 reanalysis. \textbf{b)} Hourly area-maximum (left) and area-mean (right) temperature over the heatwave core region (magenta square in \textbf{a}). NeuralGCM-ENS and NeuralGCM-Nudge results are shown in blue and pink, respectively. Solid yellow lines represent ``E3SMwCurrentLand" and dashed yellow lines represent ``E3SMwFutureLand". The black dashed line displays ERA5-Warm. As in Fig.1, daily hottest \textbf{c)} area-max temperature and \textbf{d)} area-mean temperature are examined over the heatwave core area. The histograms of ensemble involution outcomes from NeuralGCM are approximated via a Normal distribution (solid red line). The pink vertical line corresponds to the NeuralGCM-Nudge temperature simulation. The solid yellow vertical line denotes the E3SM simulation with current land conditions, while the dashed yellow line represents simulations under future land conditions. The dashed black vertical line marks ERA5-Warm. Similar as in Fig.1, the two-sided p-values from t-tests are presented for NGCM-Nudge, E3SMwCurrentLand, E3SMwFutureLand (italicized), and ERA5-Warm. These values represent the probability of obtaining temperatures as or more extreme relative to the NGCM-ENS distribution, and are color-coded for clarity. 
  }
  \label{fig:delta-PNAS}
\end{figure}

The mid-century (2050) PGW simulation results are illustrated in Fig.\ref{fig:delta-PNAS}. Similar as the hindcast case, NeuralGCM-Nudge yields the most constrained dynamics, while E3SM and NeuralGCM-ENS perform similarly, with E3SM showing better skill at 500 hPa. When examining the warming patterns, we first note that the storyline simulations generated by NeuralGCM closely match the E3SM benchmark when the land is initialized using the atmospheric forcings from the current climate (E3SMwCurrentLand). In contrast, E3SM projects substantially warmer conditions when land is spun-up with the projected future atmospheric forcings (E3SMwFutureLand; Fig.\ref{fig:delta-PNAS}\textbf{a}). This result is consistent with previous studies showing that soil moisture had a limited influence on the 2021 heatwave \cite{conrick2023influence}, but is expected to have a stronger impact in a future warmer climate \cite{bartusek20222021, bercos2022anthropogenic}. Because NeuralGCM is trained exclusively on ERA5 data and lacks a land component, it implicitly learns land–atmosphere interactions limited to the current climate. As a result, its warming projections are weaker compared to the E3SMwFutureLand results, but closely resemble the E3SM PGW simulations from E3SMwCurrentLand. 

In the following section, we continue to compare NeuralGCM with E3SMwCurrentLand, as both are based on land-air interactions under current climate conditions. A conceptual reference case is also considered where the climate change signal is added directly to ERA5 reanalysis (``ERA5-Warm"), allowing us to examine how well NeuralGCM carries forward initial warming perturbations. NeuralGCM (both -ENS and -Nudge) simulates lower temperature projections over eastern Washington than E3SMwCurrentLand and ERA5-Warm (Fig.\ref{fig:delta-PNAS}\textbf{a}), a feature already noted when comparing NeuralGCM and E3SM hindcast patterns for the historical period in Fig.\ref{fig:hindcast}. Over the core region, the temporal evolution of area-maximum and area-mean temperature in NeuralGCM-ENS aligns relatively well with the E3SMwCurrentLand and ERA5-Warm, while NeuralGCM-Nudge projects a relatively lower temperature (Fig.\ref{fig:delta-PNAS}\textbf{b}), as already noted for the historical period. Despite this systematic offset in both hindcast and PGW simulations, NeuralGCM-Nudge exhibits a more pronounced net warming between the mid-century and the historical periods, characterized by greater increases in both area-maximum and area-mean temperature time series. This warming remains of similar magnitude to the derived climate change signal represented in the ERA5-Warm case (Fig.S3). Overall, NeuralGCM shows minimal spatial and temporal differences relative to E3SMwCurrentLand and ERA5-Warm, indicating that it generates reasonable and plausible temperature projections when compared to physical-based references, provided that land-air interactions are kept under current climate conditions. The comparison with ERA5-Warm further suggests that NeuralGCM effectively preserves the initial temperature perturbation while simulating the event’s evolution similarly to the historical case. 

We now examine the results of E3SM and NeuralGCM-Nudge, comparing them to the distribution generated via NeuralGCM-ENS. Similar to the hindcast case, both NeuralGCM-Nudge and E3SMwCurrentLand fall near the center of the distribution, with p-values $>0.05$. This suggests that they cannot be rejected as sampled from NeuralGCM-ENS distribution. In other words, their simulations are statistically compatible to NeuralGCM-ENS distribution and they fall within the spread of NeuralGCM-ENS in the mid-century scenario. As anticipated, the E3SMwFutureLand cases show much higher temperature projections that fall outside the NeuralGCM-ENS distribution spread (with associated statistically significant p-values falling below the selected 0.05 level). 

As a final evaluation, we compare the present-day (distributions in Fig.\ref{fig:hindcast}\textbf{d},\textbf{e}) and future (distributions in Fig.\ref{fig:delta-PNAS}\textbf{c},\textbf{d}) NeuralGCM-ENS ensembles day by day over the whole heatwave duration (as shown in Fig.S4). In each day, the distributions are found to be statistically distinct at a 5\% significance level (Kolmogorov-Smirnov test), driven by their mean difference (t-test), while maintaining similar variances (F-test; details in Table S1). These features demonstrate that NeuralGCM is capable of generating physically plausible ensembles in the current PGW settings without regressing to its training climatology, while preserving a stable stochastic representation across different climate conditions. Both characteristics are essential for ensuring reliable performance in PGW simulations.

To our knowledge, this study is the first evaluation of PGW simulations purely conducted with NeuralGCM. Our proposed evaluation framework can generalize to other extreme temperature- or precipitation-driven events \cite{yuval2024neural} and AI-based weather models. Beyond PGW simulations, AI-based models can also be used for event attribution analyses, where the climate change delta (i.e., historical minus preindustrial conditions) is subtracted \cite{jimenez2024ai, leach2024heatwave}. However, a comprehensive evaluation would help build confidence in these results \cite{Ullrich2025}, as AI models may still drift towards their training climatology \cite{rackow2024robustness}. 
In addition to simulating the whole event, a pioneering study successfully applied machine learning to event attribution \cite{trok2024machine}, focusing on specific extreme event indices. Nonetheless, PGW simulations differ by offering detailed insights into the spatial distribution and temporal evolution of the targeted event. As a global model, NeuralGCM can also be applied to diverse events without the need of re-training. 

As our selected reference model, E3SM is not expected to be perfect; different climate models may produce varying outcomes. For example, while soil moisture–atmosphere coupling generally amplifies warming, its strength varies significantly across models \cite{qiao2023soil}, and heatwave intensification driven by land–atmosphere interactions shows inter-model variations \cite{cai2024pronounced}. Constructing a comprehensive set of PGW simulations across all major climate models is currently impractical, as it demands diverse technical expertise and substantial computational resources. Such an effort would require multi-institutional collaboration and standardized protocols — ideally within a model intercomparison project — to ensure fair comparisons, which is beyond the scope of this study. While the strength of land–atmosphere feedbacks varies across regions and models, the E3SM simulation we used in this study serves as a reasonable reference within CMIP6 models, illustrating one plausible pathway of land-atmosphere interaction under PGW settings, rather than a definitive or model-independent outcome. That said, given that NeuralGCM lacks an interactive land component, its behavior is assumed to align more closely with models that exhibit weak or negligible land-atmosphere coupling.

We emphasize that NeuralGCM is an atmospheric-only model not originally designed for PGW analyses. Our study, hence, pushes NeuralGCM to the boundaries of its abilities. Compared to E3SM, it offers significant scalability advantages, with a 2-minute simulation time per ensemble member on a single NVIDIA A-100 GPU and parallelization is straightforward with multi-GPU machines. In comparison, E3SM takes 45 minutes on eight Intel-Xeon-CLX-8276L CPUs. This computational efficiency, combined with their ability to handle large ensembles, positions AI-based models as valuable tools for climate research. Despite their promising features, challenges remain, particularly with biases linked to land-atmosphere interactions. Improving NeuralGCM for storyline analyses will require integrating land-atmosphere dynamics or developing fully coupled AI-based models, enhancing accuracy in regions where local feedbacks are crucial. Addressing these limitations could enable AI-based weather models to complement traditional approaches, transforming how we predict and analyze extreme events.

\clearpage

\section*{Methods}
\subsection*{Model Configurations}

NeuralGCM is a hybrid AI-based global climate model \cite{kochkov2024neural} that integrates differentiable dynamical cores and learnable physical schemes. It predicts atmospheric conditions at the next time step based on initial conditions and forcing variables. By employing an auto-regressive approach, NeuralGCM can be applied to a variety of tasks, ranging from medium-range weather forecasts to seasonal and long-term climate projections\cite{zhang2025seasonal}. The model is available in two versions: deterministic and stochastic. The deterministic version produces a consistent output for a given initial condition and is available at three horizontal resolutions ($0.8^{\circ}$, $1.4^{\circ}$, and $2.8^{\circ}$); for this study, we selected the $1.4^{\circ}$ version. In contrast, the stochastic version supports ensemble generation through the use of a random key and is currently available only at $1.4^{\circ}$ resolution. We apply nudging to the deterministic model and ensemble involution to the stochastic model. 

The Energy Exascale Earth System Model (E3SM) is a state-of-the-art, fully coupled global climate model \cite{golaz2022doe}, encompassing multiple components such as atmosphere, land, ocean, sea-ice, and rivers. E3SM offers a flexible range of resolutions, from 1 degree (low-resolution) down to 3 km (convection-permitting) for atmosphere component \cite{Donahue2024}, and has regionally refined configuration \cite{Zhang_2024}. In this study, we employed its low-resolution version to ensure consistency in spatial resolution when compared with NeuralGCM. We used hindcast configuration (reanalysis atmosphere and land initial conditions as well as prescribing sea surface temperature and sea ice), a configuration that makes climate simulations run in weather forecasting mode relying on accurate initial conditions rather than long-term forcings. The simulation was performed deterministically with a single realization. The results are regridded to the NeuralGCM $1.4^{\circ}$ with conservative regridding method. 

It is notable that NeuralGCM operates at a much coarser spatial resolution compared to other AI-based weather models, such as Pangu and FourcastNet \cite{bi2023accurate, bonev2023spherical}, which achieve resolutions as fine as $0.25^{\circ}$. Despite this limitation, we selected NeuralGCM because it shows some potential to extrapolate long-term simulations to warmer climate conditions in atmospheric-only settings (e.g., +2k sea surface temperature simulations\cite{kochkov2024neural}). This makes it particularly suitable for investigating thermodynamic responses. However, it is important to note that in previously used cases\cite{kochkov2024neural} extrapolation did not account for land–atmosphere interactions, which may contribute significantly to regional extremes. In contrast, other models have been shown to regress toward training climatology under warming conditions\cite{rackow2024robustness}. To ensure a fair comparison, we opted for E3SM rather than commonly-used regional climate models, e.g., Weather Research Forecast model (WRF), which can resolve finer details. It is promising to explore the use of other AI-based models to downscale NeuralGCM’s results, which helps refine spatial resolution and address the resolution gap, leading to better representations of localized extreme events. In such cases, regional climate models or regionally refined GCMs would become more appropriate for comparison, providing simulations at finer scales. Nonetheless, our work aims to demonstrate the capability of AI-based weather models for storyline analysis studies, and we anticipate broader applications of these models in future research. 

The initial conditions for both NeuralGCM and E3SM are derived from ERA5 reanalysis data. The variables are regridded to a $1.4^{\circ}$ horizontal resolution for NeuralGCM. The preprocessing of ERA5 for E3SM follows a previous study \cite{Zhang_2024}. 
All simulations begin at 00Z on June 26, 2021, and conclude at 00Z on July 2, 2021. The forcing variables (boundary conditions) include sea surface temperature (SST) and sea ice concentration for NeuralGCM and E3SM, both sourced from the ERA5 dataset. The storyline simulations are conducted in a similar way, but with the climate change signal added to the initial conditions and bottom boundary conditions. Changes (deltas) in air temperature, SST and sea ice concentration are derived from the Community Earth System Model 2 (CESM2) large ensemble simulations, following the Shared Socioeconomic Pathway 3-70 (SSP370) scenario to project future climate conditions \cite{rodgers2021ubiquity}. Specific humidly in 2050 is calculated by assuming a fixed relative humidity. The deltas are calculated as the differences between the July ensemble means for 2050 and 2021. A reference case (``ERA5-Warm") is constructed by adding the climate change signal directly to the ERA5 reanalysis data. Because the signal is derived from monthly and ensemble means, it lacks daily or sub-daily variability, as displayed in Fig.S3.

The land component of E3SM is spun-up using five years of atmospheric forcings in both hindcast and PGW settings to provide stable land states that are consistent with the corresponding climate conditions. Since NeuralGCM is an atmospheric-only model, E3SM land conditions are forced using the current climate for both hindcast and projections (E3SMwCurrentLand) to ensure a fair comparison. Additionally, we apply the PGW deltas to the atmospheric conditions and use the projected climate to spin-up the land (E3SMwFutureLand). This approach represents the simulations with the land-air interactions in a warmer climate.

\subsection*{Nudging}

Nudging is a technique that adjusts simulated variables toward a reference dataset by applying an additional tendency term, with the conceptual goal to gently align the simulations with observed dynamical conditions. It can be considered a simplified form of data assimilation, which is commonly used in weather forecasting to integrate multiple data sources (e.g., satellite and in-situ observations). Unlike data assimilation methods, which involves multiple observations, nudging typically relies on a single reference dataset to gradually constrain the model toward observed or reanalysis conditions. 
To simulate the extreme event under the same dynamical conditions, nudging is used for E3SM. Global simulated horizontal and meridional wind fields are nudged to ERA5 at every physics timestep (15 minutes) for both the hindcast and storyline simulations. The $0.25^{\circ}$ ERA5 pressure level data are interpolated in advance to the E3SM model grid (1 degree, 72 hybrid vertical levels). The nudging relaxation timescale ($t_\tau$) is set to 3 hours, which determines the nudging tendency strength. We refer the readers to the reference study \cite{Zhang_2024} for detailed information regarding nudging in E3SM. 

The current NeuralGCM codebase (v1.1.2) lacks an explicit capability to perform nudging. Nevertheless, developing a nudging algorithm for NeuralGCM, analogous to those employed in traditional process-based climate models, holds significant potential. In this study, we implemented a straightforward ``nudging" approach, where circulations (i.e., wind fields) are adjusted by adding a tendency term to the simulated winds. This tendency term is calculated as:
\begin{align}
    X_{\text{nudge}}^t &= \frac{X_{\text{ref}}^{t}-X_{\text{sim}}^{t}}{t_\tau}\\
    X_{\text{adjust}}^t &= X_{\text{sim}}^{t}+X_{\text{nudge}}^t\times \Delta t
\end{align}
Here, $X_{ref}^{t}$ represents the reference variables (ERA5 wind fields in this study) at time step $t$, and $X_{sim}^{t}$ denotes the corresponding simulated variable at time step $t$. The nudging relaxation timescale ($t_\tau$) is set to 3 hours to keep consistent with E3SM. Similar as E3SM, the nudging is conducted at every physics timestep (i.e., 1 hour for NeuralGCM). The computed nudging tendency is scaled by the model's timestep ($\Delta t=$1 hour) and subsequently added to the simulated variables. These adjusted variables are then employed for the next timestep iteration. It is important to note that the proposed nudging method is not intended to be definitive or optimal. As the state variables are adjusted at each timestep, this method is analogous to re-initialization, and we noticed an initialization shock with the stochastic version of the NeuralGCM model (as shown in Fig.S6), which fades after six hours but persists in the case of re-initialization. Consequently, this approach is only applicable to the deterministic version of the model, producing a single outcome rather than a probabilistic distribution of results. For a better approach to avoid re-initialization and a comprehensive evaluation of nudging techniques, a dedicated and independent study would be required. As demonstrated in prior research, different nudging implementations can lead to varying outcomes \cite{zhang2022further}. 

\subsection*{Ensemble Involution}

In addition to explicitly implementing nudging, we also employed ensembles to mimic the behavior of nudging, providing an alternative way to constrain dynamical conditions. 
Within the context of AI-based models, there have been several studies and approaches focusing on creating large and huge ensembles to better simulate extreme events \cite{mahesh2024huge}. While effective for extreme weather prediction, such large ensembles are less suitable for storyline analyses, as their dynamical conditions would diverge from the target event. Here, we introduce \textit{ensemble involution}, a novel approach for AI-based weather models for climate applications. Instead of expanding the ensemble size, ensemble involution begins with a large ensemble and \textit{reduces} its size by iteratively selecting the most skillful members. Though counterintuitive, this method improves the prediction of extreme events by concentrating on the most reliable simulation trajectories, thereby enhancing the representation of their dynamical evolution. 

The ensemble involution process proceeds through the following steps: Initially, a large ensemble of simulations is generated with the stochastic model of NeuralGCM, consisting of a significant number of members (1000 in our case). At each time interval (every 24 hours in our study), the instantaneous predicted representative dynamic conditions over a specific domain—specifically zonal and meridional wind at 850 hPa and 500 hPa—are evaluated against the hourly target fields derived from ERA5 reanalysis data. We assessed the prediction skill using the coefficient of determination ($R^2$), defined as:
\begin{align}
    R^2 = 1-\frac{\sum_i(\hat{y_i}-y_i)^2}{\sum_i(\overline{y}-y_i)^2}
\end{align}
where $y$ represents the reference variable, $\hat{y}$ denotes the predicted values, $\overline{y}$ is the reference variable mean over the domain and $i$ indicates the spatial grid points. A higher $R^2$ indicates better prediction skill while an $R^2$ of zero corresponds to the skill level achieved by predicting the mean value of the variable across the domain. 
Ensemble members that exhibit a negative skill score in any dynamic field are eliminated, and from the remaining members, the top 80\% are selected based on their average performance, which is the average R square score from the four selected variables. This iterative refinement is repeated at each time step, progressively narrowing the ensemble spread until the final prediction date is reached (Fig.S5 illustrates the skill score and ensemble size evolutions through this process). Unlike traditional model evaluations at target time slices, which risk to fully miss the whole trajectory of an extreme event, ensemble involution retains only the most skillful members that consistently capture the dynamical evolution of the atmospheric circulation. 

In the current configuration, wind fields at 500 hPa and 850 hPa from $30^\circ N$ to $60^\circ N$ and $140^\circ W$ to $110^\circ W$ are used as representative variables, and ensemble involution is conducted at 24-hour intervals. This reduces the initial 1000 ensemble members to 86 for the hindcast case (Fig.\ref{fig:hindcast}) and 54 for the future projections (Fig.\ref{fig:delta-PNAS}). We conducted sensitivity tests using either a larger spatial domain (from $10^\circ N$ to $80^\circ N$ and $160^\circ W$ to $90^\circ W$), or additional vertical levels (200 hPa, 500 hPa, 850 hPa and 925 hPa). These different configurations yield similar synoptic circulations (Fig.S7), primarily because atmospheric variables are inherently correlated across spatial and vertical scales. Even when expanding the domain or including additional vertical levels, the large-scale dynamical constraints ensure that the overall circulation patterns remain consistent. Consequently, they produce comparable temperature patterns and temporal evolutions (Fig.S7). Statistical comparisons of the hottest area-maximum and area-mean temperatures distributions across different ensemble involution configurations reveal negligible differences between them (Table S2). 

Our proposed ensemble involution diverges from other ensemble-based approaches designed for weather forecast or climate simulations in previous studies \cite{sparrow2018finding, fischer2023storylines}, which tend to increase or preserve the ensemble size by re-initializing selected members. Ensemble involution reduces ensemble size through the elimination of members that deviate from reference conditions. This approach prioritizes constraining dynamical conditions rather than enhancing predictability, thereby conceptually mimicking the effects of nudging. It offers a simpler and more efficient solution tailored to this objective.
In addition, unlike re-initialization approaches that constrained ensemble members based on target variables \cite{sparrow2018finding, fischer2023storylines}, ensemble involution evaluates members solely using wind fields, without requiring optimal performance on temperature. This makes it particularly effective for storyline simulations, as it facilitates meaningful comparisons of temperature responses. Furthermore, ensemble involution is well-suited for AI-based models, which are computationally efficient and naturally incorporate ensemble capabilities, particularly when trained with stochastic methods (such as the stochastic model of NeuralGCM). Moreover, instability is a significant challenge for data-drive models, as demonstrated in a recent study \cite{slivinski2024assimilating}, as well as the initialization shock in our nudging implementation (Fig.S6). Ensemble involution circumvents this issue and eliminates concerns about instability, thereby enhancing its practical applicability.

\subsection*{Statistical Tests}
The NeuralGCM ensembles (NGCM-ENS) are examined via multiple statistical tests. First, we fit a Normal distribution to the distribution of ensemble outcomes (e.g., area-mean or area-maximum temperatures). To examine whether the deterministic results (i.e., ERA5, NeuralGCM-Nudge and E3SM) can be considered as coming from the NeuralGCM-ENS distributions (i.e., NeuralGCM-ENS can statistically represent these results), a t-statistic is calculated with the following equation:
\begin{align}
    t = \frac{x-\mu}{s/\sqrt{n}}
\end{align}
where $x$ denotes the deterministic results. The mean and variance of NeualGCM-ENS are denoted as $\mu$ and $s$ respectively, and $n$ represents the ensemble size. The two-sided p-values are consequently determined based on the Normal distribution and the t-statistics:
\begin{align}
    p = (1-\text{Cdf}(|t|))\times 2
\end{align}
with $\text{Cdf}$ denoting the cumulative distribution function. A p-value greater than 0.05 ($\alpha$) indicates that the deterministic result can be considered statistically consistent with the distribution and could plausibly be drawn from it at the $\alpha$ significance level. 

When comparing NeuralGCM-ENS distributions between the current and future climates, we employ the Kolmogorov-Smirnov (K-S) test to assess whether the distributions differ, the t-test to compare their means, and the F-test to evaluate differences in variance. The corresponding statistics ($D, t$ and $F$) are defined as follows:
\begin{align}
    D_{m, n} &= \text{sup}_x|\text{Cdf}_{\text{current}, m}(x)-\text{Cdf}_{\text{future}, n}(x)|\\
    t &=\frac{\mu_{\text{current}}-\mu_{\text{future}}}{\sqrt{s_{\text{current}}/m+s_{\text{future}}/n}}\\
    F&=\frac{s_{current}}{s_{future}}
\end{align}
where $m$ and $n$ denote the ensemble sizes for the current and future NeuralGCM-ENS, respectively. The empirical cumulative distribution functions are represented as $\text{Cdf}$, and $sup$ denotes the supremum difference function. A p-value larger than 0.05 indicates that, at the 0.05 significance level, there is insufficient evidence to reject the null hypothesis that the distributions are statistically indistinguishable (K-S test), have the same mean (t-test), and exhibit equal variance (F-test).

\bibliography{ref}

\section*{Acknowledgements}

This work was performed under the auspices of the U.S. Department of Energy by Lawrence Livermore National Laboratory under Contract DE-AC52-07NA27344. The manuscript is released under LLNL-JRNL-870890. S.D., C.B., and G.P. are funded by PCMDI Science Focus Area. The study has been initiated using LLNL-LDRD Project 22-ERD-052 (PI: P. Caldwell) PlusUp Initiative led by C.B. S.D. also received support from ISCP-23PLS021-Climate for technical work related to installing and running NeuralGCM in test cases. J.Z. is funded by LLNL-LDRD Project 22-SI-008. The authors would like to acknowledge the help from Dr. Peter Bogenschutz and Dr. Chris Golaz on E3SM simulations. 

\section*{Author contributions statement}

S.D., J.Z., C.B., and G.P. designed the experiments and formulated the article structure. S.D. and J.Z. conducted the experiments and analyzed the results. All authors reviewed the manuscript.

\section*{Additional information}

\textbf{Code availability} The NeuralGCM model is publicly accessible through its GitHub repository: \nolinkurl{https://github.com/google-research/neuralgcm}. Our simulation and analysis code is available in the GitHub repository: \nolinkurl{https://github.com/ShihengDuan/NeuralStoryHeat}. 

\textbf{Data availability} ERA5 is publicly available at Copernicus Climate Change Service: \nolinkurl{https://cds.climate.copernicus.eu}. Our simulation results are available in the Zenodo repository: \nolinkurl{10.5281/zenodo.13937375}. 

\textbf{Competing interests} The authors declare no competing interests.




\end{document}